\documentclass[twoside,twocolumn,9pt]{article}
\usepackage{extsizes}
\usepackage[super,sort&compress,comma]{natbib} 
\usepackage[version=3]{mhchem}
\usepackage[left=1.5cm, right=1.5cm, top=1.785cm, bottom=2.0cm]{geometry}
\usepackage{balance}
\usepackage{mathptmx}
\usepackage{sectsty}
\usepackage{graphicx} 
\usepackage{lastpage}
\usepackage[format=plain,justification=justified,singlelinecheck=false,font={stretch=1.125,small,sf},labelfont=bf,labelsep=space]{caption}
\usepackage{float}
\usepackage{fancyhdr}
\usepackage{fnpos}
\usepackage[english]{babel}
\addto{\captionsenglish}{%
  
}
\usepackage{array}
\usepackage{droidsans}
\usepackage{charter}
\usepackage[T1]{fontenc}
\usepackage[usenames,dvipsnames]{xcolor}
\usepackage{setspace}
\usepackage[compact]{titlesec}
\usepackage[hidelinks]{hyperref}

\usepackage{amsmath}
\usepackage[utf8x]{inputenc}

\usepackage{epstopdf}

\usepackage{xcolor}
\usepackage{ulem}
\newcommand\hl{\bgroup\markoverwith
  {\textcolor{white}{\rule[-.5ex]{2pt}{2.5ex}}}\ULon}

\definecolor{cream}{RGB}{222,217,201}

\begin{document}

\pagestyle{fancy}
\thispagestyle{plain}
\fancypagestyle{plain}{
\renewcommand{\headrulewidth}{0pt}
}

\makeFNbottom
\makeatletter
\renewcommand\LARGE{\@setfontsize\LARGE{15pt}{17}}
\renewcommand\Large{\@setfontsize\Large{12pt}{14}}
\renewcommand\large{\@setfontsize\large{10pt}{12}}
\renewcommand\footnotesize{\@setfontsize\footnotesize{7pt}{10}}
\makeatother

\renewcommand{\thefootnote}{\fnsymbol{footnote}}
\renewcommand\footnoterule{\vspace*{1pt}%
\color{cream}\hrule width 3.5in height 0.4pt \color{black}\vspace*{5pt}} 
\setcounter{secnumdepth}{5}

\makeatletter 
\renewcommand\@biblabel[1]{#1}            
\renewcommand\@makefntext[1]%
{\noindent\makebox[0pt][r]{\@thefnmark\,}#1}
\makeatother 
\renewcommand{\figurename}{\small{Fig.}~}
\sectionfont{\sffamily\Large}
\subsectionfont{\normalsize}
\subsubsectionfont{\bf}
\setstretch{1.125} 
\setlength{\skip\footins}{0.8cm}
\setlength{\footnotesep}{0.25cm}
\setlength{\jot}{10pt}
\titlespacing*{\section}{0pt}{4pt}{4pt}
\titlespacing*{\subsection}{0pt}{15pt}{1pt}

\fancyfoot{}
\fancyfoot[RO]{\footnotesize{\sffamily{1--\pageref{LastPage} ~\textbar  \hspace{2pt}\thepage}}}
\fancyfoot[LE]{\footnotesize{\sffamily{\thepage~\textbar\hspace{4.65cm} 1--\pageref{LastPage}}}}
\fancyhead{}
\renewcommand{\headrulewidth}{0pt} 
\renewcommand{\footrulewidth}{0pt}
\setlength{\arrayrulewidth}{1pt}
\setlength{\columnsep}{6.5mm}
\setlength\bibsep{1pt}

\makeatletter 
\newlength{\figrulesep} 
\setlength{\figrulesep}{0.5\textfloatsep} 

\newcommand{\topfigrule}{\vspace*{-1pt}%
\noindent{\color{cream}\rule[-\figrulesep]{\columnwidth}{1.5pt}} }

\newcommand{\botfigrule}{\vspace*{-2pt}%
\noindent{\color{cream}\rule[\figrulesep]{\columnwidth}{1.5pt}} }

\newcommand{\dblfigrule}{\vspace*{-1pt}%
\noindent{\color{cream}\rule[-\figrulesep]{\textwidth}{1.5pt}} }

\makeatother

\twocolumn[
\begin{@twocolumnfalse}
\begin{tabular}{m{4.5cm} p{13.5cm} }

& 
\noindent\LARGE{\textbf{Band gap opening from displacive instabilities in layered covalent-organic frameworks 
$^\dag$}} \\
\vspace{0.3cm} & \vspace{0.3cm} \\

& \noindent\large{Ju Huang,\textit{$^{a}$} Matthias J. Golomb,\textit{$^{a}$} Seán R. Kavanagh,\textit{$^{a}$$^{,}$$^{b}$} Kasper Tolborg,\textit{$^{a}$}  Alex M. Ganose,\textit{$^{a}$}  Aron Walsh$^{\ast}$\textit{$^{a}$$^{,}$$^{c}$} } \\

\\
& \noindent \normalsize{Covalent organic frameworks (COFs) offer a high degree of chemical and structural flexibility. There is a large family of COFs built from 2D sheets that are stacked to form extended crystals. While it has been common to represent the stacking as eclipsed with one repeating layer (``AA''), there is growing evidence that a more diverse range of stacking sequences is accessible. Herein, we report a computational study using density functional theory of layer stacking in two prototypical COFs, Tp-Azo and DAAQ-TFP, which have shown high performance as Li-ion battery electrodes. We find a striking preference for slipped structures with horizontal offsets between layers ranging from 1.7 \r{A} to 3.5 \r{A} in a potential energy minimum that forms a low energy ring. The associated symmetry breaking results in a pronounced change in the underlying electronic structure. A band gap opening of 0.8 -- 1.4 eV is found due to modifications of the underlying valence and conduction band dispersion as explained from changes in the $\pi$ orbital overlap. The implications for the screening and selection of COF for energy applications are discussed. 
}
\end{tabular}

\end{@twocolumnfalse}\vspace{0.6cm}
]

\renewcommand*\rmdefault{bch}\normalfont\upshape
\rmfamily
\section*{}
\vspace{-1cm}


\footnotetext{\textit{$^{a}$~Thomas Young Centre and Department of Materials, Imperial College London, Exhibition Road, London SW7 2AZ, UK}}
\footnotetext{\textit{$^{b}$~Thomas Young Centre and Department of Chemistry, University College London, 20 Gordon Street, London WC1H 0AJ, UK}}
\footnotetext{\textit{$^{c}$~Department of Materials Science and Engineering, Yonsei University, Seoul 03722, Korea }}






\section{Introduction}
Covalent organic frameworks (COFs) are porous organic materials that can adopt various topologies, using linkers to form periodic skeletons and ordered nanopores in two and three dimensions (2D and 3D).\cite{cote2005porous, diercks2017atom,huang2016covalent,geng2020covalent,lohse2018covalent} 
In layered COFs, which were reported in 2005\cite{cote2005porous}, the organic units are linked by strong in-plane covalent bonds to form 2D sheets, which can then be stacked into crystalline structures.\cite{winkler2021understanding} $\pi$-$\pi$ interactions between the stacked aromatic building blocks strongly affect both the atomic and electronic structure, determining the stacking sequence, band dispersion and band gap energy.\cite{kuc2020proximity} Proximity effects from the repulsive electrostatic interactions between hydrogen and the $\pi$ system of adjacent aromatic rings cause the fully eclipsed stacking of 2D COFs to be unfavourable.\cite{kuc2020proximity,hunter1990nature} In 2007, 3D COFs were successfully synthesized using alternative building-unit geometries which were strongly connected by covalent bonds.\cite{el2007designed} Both 2D and 3D COFs exhibit the advantages of flexible and customizable crystal structures, high chemical and thermal stability, and high porosity – making them promising candidates for applications such as energy storage\cite{liu2020porous,shi2020nitrogen,rojaee2020two,deblase2015rapid,kong2021redox,sun2020covalent,song2021covalent,lei2018boosting,ball2020triazine,luo2018microporous,wu20202d,yao2020two,an2021designs}, ion and molecule separation,\cite{li2020laminated, dey2020nanoparticle,tong2017exploring} optoelectronics\cite{li2018tuneable,lv2018direct} and catalysis\cite{bi2019two,liu2022covalent,xu2015stable,fu2020stable,zhu2020efficient}.    

\begin{figure*}[t!]
    \centering
    \includegraphics[width=0.8\textwidth]{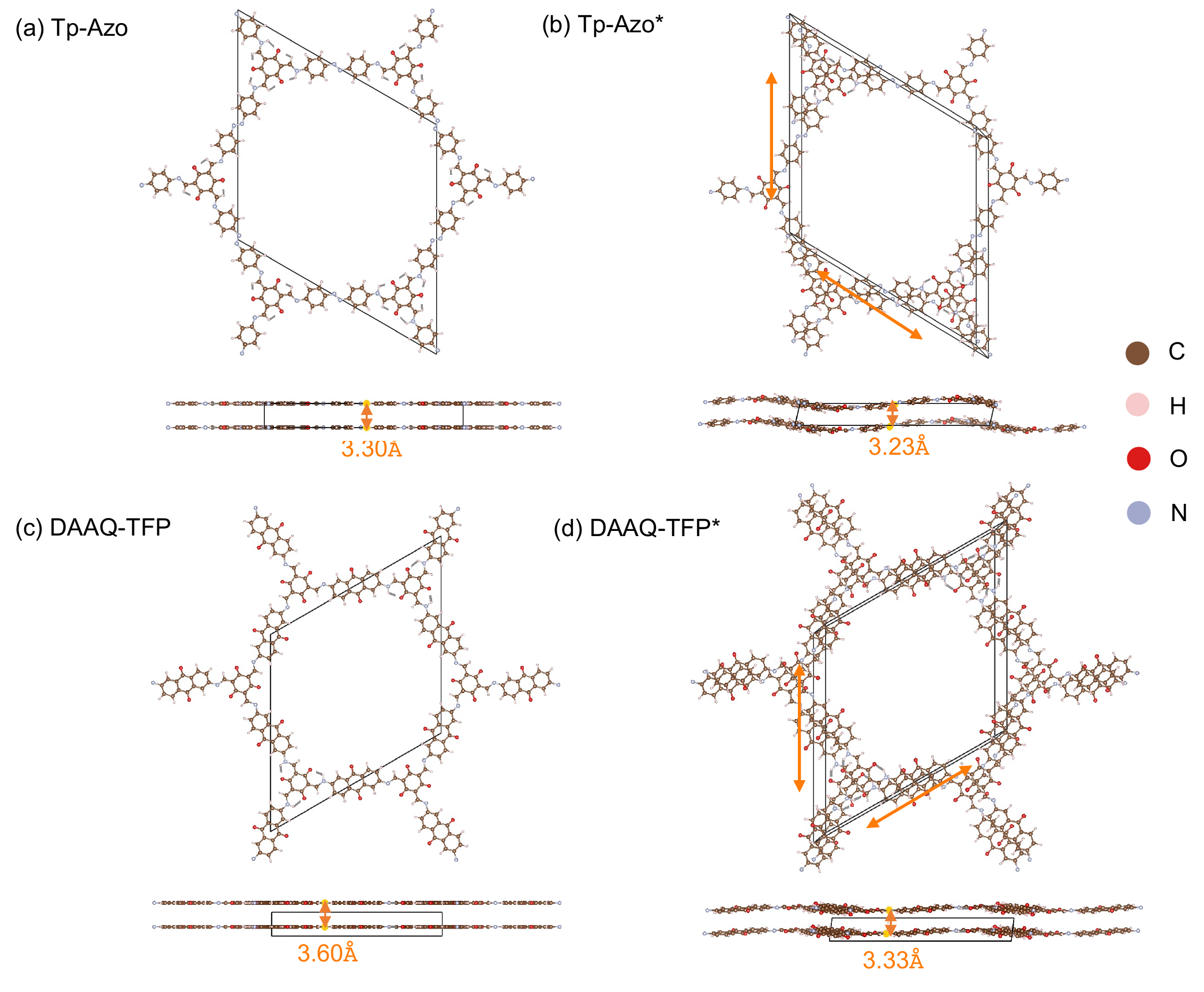}
    \caption{\textbf{(a)} Planar experimentally-reported crystal structure and \textbf{(b)} wavy relaxed structure of Tp-Azo. \textbf{(c)} Planar experimentally-reported structure and \textbf{(d)} wavy relaxed structure of DAAQ-TFP COF. The upper figures are viewed looking down the $c$ axis, and the lower figures are viewed along the $ab$ plane. ``$*$'' refers to the relaxed crystal structures. The arrows in (b) and (d) are the displacement directions along the $a$ and $b$ unit cell vectors. }
    \label{fig:1}
\end{figure*}

Various forms of disorder exist in experimentally-synthesized COFs, such as bond breakage, pore collapse and stacking faults. Such imperfections can significantly affect the properties of 2D COFs, causing loss of crystallinity,  porosity, and conductivity.\cite{haase2017tuning, koo2012classification, spitler20112d, putz2020total,emmerling2021interlayer} In particular, the interlayer stacking modes of 2D aromatic COFs play a critical role in determining their properties. The stacking behaviour of COFs is not thoroughly understood, however, due to difficulties in experimental characterisation of the dynamic, low-crystallinity materials. For instance, powder X-ray diffraction (XRD) can only detect the existence of crystalline domains, making the extraction of accurate results difficult in the presence of low long-range order and sizeable thermal dynamics.\cite{kuc2020proximity,lukose2010reticular} As such, XRD measurements struggle to quantitatively distinguish crystalline structures from other similar aggregated structures, as a result of peak broadening in the diffraction pattern.\cite{kang2022aggregated,zhou2010structural, putz2020total, winkler2021understanding} To achieve greater resolution of COF layer stacking, Kang et. al\cite{kang2022aggregated} recently used $^{13}\textrm{C}$ solid-state nuclear magnetic resonance (ssNMR) to distinguish different aggregated structures by studying the interactions between atoms and chemical groups from adjacent layers.

Five different stacking modes in 2D COFs have been reported: eclipsed, inclined, zigzag, staggered and random stacked.\cite{putz2020total,mahringer2020taking} The eclipsed stacking (AA) corresponds to zero horizontal (coplanar) offset between neighbouring layers in the $ab$ plane, which has the highest symmetry and is the most often reported in experimental works. The inclined stacking (AA\textquoteright) corresponds to a constant, collinear offset between neighbouring layers. This stacking mode was observed using powder XRD and transmission electron microscopy (TEM) in SIOC-COF-8 and SIOC-COF-9.\cite{fan2017case} Zigzag stacking (AB) corresponds to an alternating offset direction between layers but it still retains high porosity in the stacking sequence. Staggered stacking is a special type of AB stacking, whereby the offset between layers is sufficient to make one layer's skeleton centered directly above the other pore, e.g.~a horizontal offset halfway along the $ab$ unit cell diagonal. This large offset between layers would reduce the porosity completely in the structures.\cite{putz2020total} These four stacking modes can be combined to form a random stacking sequence\cite{mahringer2020taking}, which is difficult to characterize experimentally or computationally due to limitations in equipment precision and computational demand. 

Several studies have focused on stacking modes and their effect on properties for various 2D COFs.\cite{putz2020total,haase2017tuning,koo2012classification,spitler20112d,emmerling2021interlayer} It has been found that the AA stacking mode is the most energetically unfavorable as a result of strong repulsive interlayer orbital interactions.\cite{lukose2010reticular, koo2012classification, haase2017tuning} Koo et al. studied the potential energy surface (PES) of 33 COFs using molecular mechanics (MM) and density functional theory (DFT) approaches, finding that COFs are preferentially stacked with 1--2 \r{A} horizontal offsets between layers.\cite{koo2012classification} It has been reported that bulk COF structures have either inclined or zigzag stacking, which are more energetically favorable than eclipsed and staggered stacking.\cite{lukose2011structure, winkler2021understanding} The simulated XRD patterns of unidirectionally slipped (AA\textquoteright) and alternating slipped (AB) modes show a better agreement with the experimental XRD pattern than the eclipsed structures.\cite{lukose2011structure,putz2020total} More precisely, in many studies\cite{martinez2021understanding,fan2017case,zhou2010structural}, the predicted diffraction patterns of inclined stacking are more consistent with experiment than other stacking modes. 

COFs of Tp-Azo\cite{chandra2014phosphoric} and DAAQ-TFP\cite{deblase2013beta}, whose experimentally reported crystal structures are shown in Fig.~\ref{fig:1}(a) and (c), have been reported with high energy capacity, good cycling performance and excellent stability as battery electrodes.\cite{zhao2020dual,an2021designs,wang2017exfoliation, deblase2013beta} It has been predicted that 30 Li$^+$ ions per unit cell can be inserted into and extracted from the porous Tp-Azo structure, using DFT simulations.\cite{zhao2020dual} DAAQ-TFP COF linked by $\beta$-ketoenamines\cite{kandambeth2012construction, chandra2013chemically} was the first COF to exhibit reversible redox behavior in energy storage systems and has the highest surface area of all COFs linked by either imines or enamines.\cite{deblase2015rapid,deblase2013beta} However, the basic structural properties, the stacking modes and the electronic structures in Tp-Azo and DAAQ-TFP COFs have not been reported. In this work, we present a theoretical study of the bulk properties and potential energy surface for stacking fault disorder of these two COFs. Furthermore, we investigate the effect of the stacking sequence on the electronic structure, rationalising the behaviour through consideration of the interlayer orbital interactions, and discuss the implications for COF material design for energy applications. 

\section{Methods}
All electronic structure calculations were performed using Kohn--Sham DFT through the all-electron ``Fritz Haber Institute ab initio molecular simulations'' \textit{FHIaims} package.\cite{blum2009ab, ren2012resolution, havu2009efficient, levchenko2015hybrid} Both the semi-local functional of Perdew-Burke-Ernzerhof revised for solids (PBEsol)\cite{perdew2008restoring} and the hybrid Heyd-Scuseria-Ernzerhof (HSE06)\cite{heyd2003hybrid} functional were used, and the Tkatchenko-Scheffler correction was implemented to account for van der Waals (vdWs) interactions between layers. The PBEsol functional was used for geometry optimisation, having been shown to predict atomic structures and energies of solid materials with good accuracy.\cite{perdew2008restoring} The HSE06 functional was used for calculations of electronic band structures, having been shown to accurately reproduce the electronic structure across a range of semiconductors.\cite{borlido2020exchan} A \textit{k}-point grid of $1 \times 1 \times 10$ was used for the geometry optimisation with $6 \times 6 \times 6$ sampling used for electronic structure analysis. 
An energy convergence criterion of 0.01 meV per unit cell was used with an atomic force tolerance of 0.01 eV/\r{A}.

The initial crystal structure parameters of Tp-Azo and DAAQ-TFP were obtained from the CoRE-COF database\cite{tong2017exploring}. These structures were firstly relaxed using the lighter Tier 1 numerical basis set, followed by a relaxation with the expanded Tier 2 basis set, before calculating the energetic and electronic properties. The well-converged conventional ``intermediate'' basis functions for each element species were used in the band structure calculations.

\begin{figure}[t!]
    \centering
    \includegraphics[width=\columnwidth]{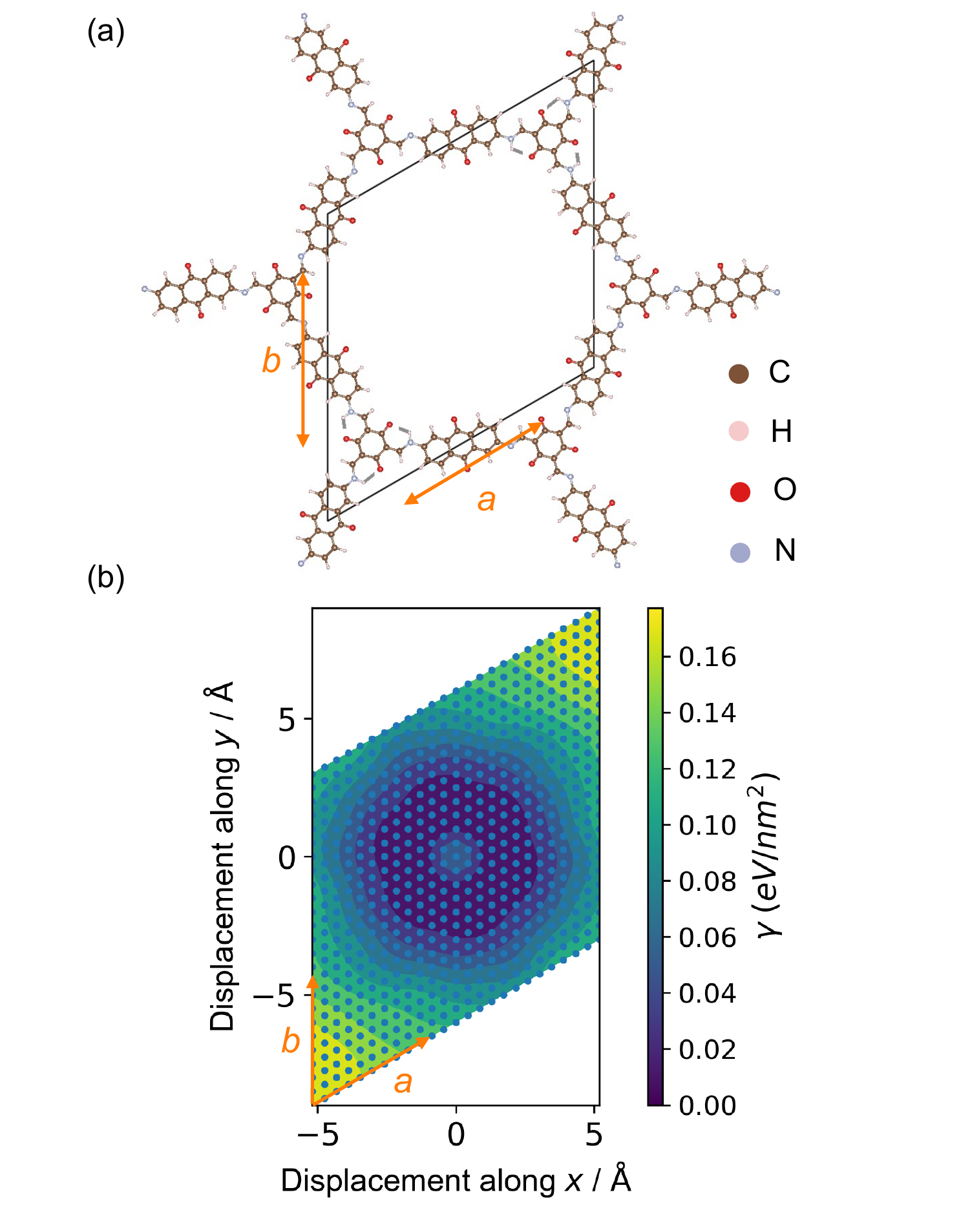}
    \caption{\textbf{(a)} The experimentally-reported crystal structure of DAAQ-TFP viewed along <0001>. \textbf{(b)} Grid-based approach for the displacement of DAAQ-TFP COF along the $a$ and $b$ sides. Here the structure within each layer is held fixed as the layers are displaced. The angle between $a$ and $b$ in \textbf{(a)} and \textbf{(b)} corresponds to $\angle\gamma$ for unit cells of relaxed Tp-Azo and DAAQ-TFP. The blue dots on \textbf{(b)} represent the locations of single-point calculations using PBEsol functional.}
    \label{fig:2}
\end{figure}

The relaxed structures were modified to study the stacking fault behaviour. Layer displacement was modelled by changing the angles of $\alpha$ and $\beta$ of the unit cell, thereby shifting the individual pseudo-hexagonal layers along the $ab$ plane to yield inclined stacking modes.\cite{haase2017tuning} Fig.~\ref{fig:2} shows the slip grid of one layer to the another layer along the $a$ and $b$ sides, with offsets of -6 \r{A} to 6 \r{A} and displacement steps of 0.5 \r{A}. The distance between adjacent layers was kept fixed to that of the relaxed structures. The energies of the displaced structures were then calculated with fixed atomic positions and with relaxed atomic positions. The relaxed displaced structures were used to study the effect of the stacking faults on the physical properties.

\section{Results and Discussion}
\subsection{Crystal structure optimisation}
The crystal structure of Tp-Azo was assigned to a hexagonal $P6/m$ space group, with eclipsed stacking of planar layers separated by a distance of 3.3 \r{A}, on the basis of powder XRD measurements (Fig.~\ref{fig:1}\textbf{(a)} and Supplementary Tab.~S1\textbf{(a)}).\cite{chandra2014phosphoric}  
Similarly, the structure of DAAQ-TFP has been assigned to a $P6/m$ space group from Pawley refinement of powder XRD patterns (Fig.~\ref{fig:1}\textbf{(c)} and Supplementary Tab.~S1\textbf{(c)}).\cite{deblase2013beta}
Upon geometry optimisation, in both cases we find both a breaking of the planarity within layers through an undulating distortion, as well as relative coplanar displacements between layers as shown in Fig.~\ref{fig:1}\textbf{(b)} and \textbf{(d)}. 
The space group symmetry lowers to P$\bar{1}$.
The interlayer distance of Tp-Azo it decreases from 3.30 \r{A} to 3.23 \r{A}, and in DAAQ-TFP it decreases from 3.60 \r{A} to 3.33 \r{A} during geometry relaxation from the reference structures. 
The layer shift of Tp-Azo along $a$ is -2.63 \r{A} and along $b$ is 2.01 \r{A}, the offset between neighbouring layers along the $ab$ plane is 2.73 \r{A}. The layer shift of DDAQ-TFP in the $ab$ plane is 2.29 \r{A}. 
The horizontal offsets of both Tp-Azo and DAAQ-TFP are higher than other COFs, which have been reported with offsets of 1-2 \r{A} between neighbouring layers.\cite{koo2012classification,haase2017tuning} 

\begin{figure*}[t!]
    \centering
    \includegraphics[width=0.8\textwidth]{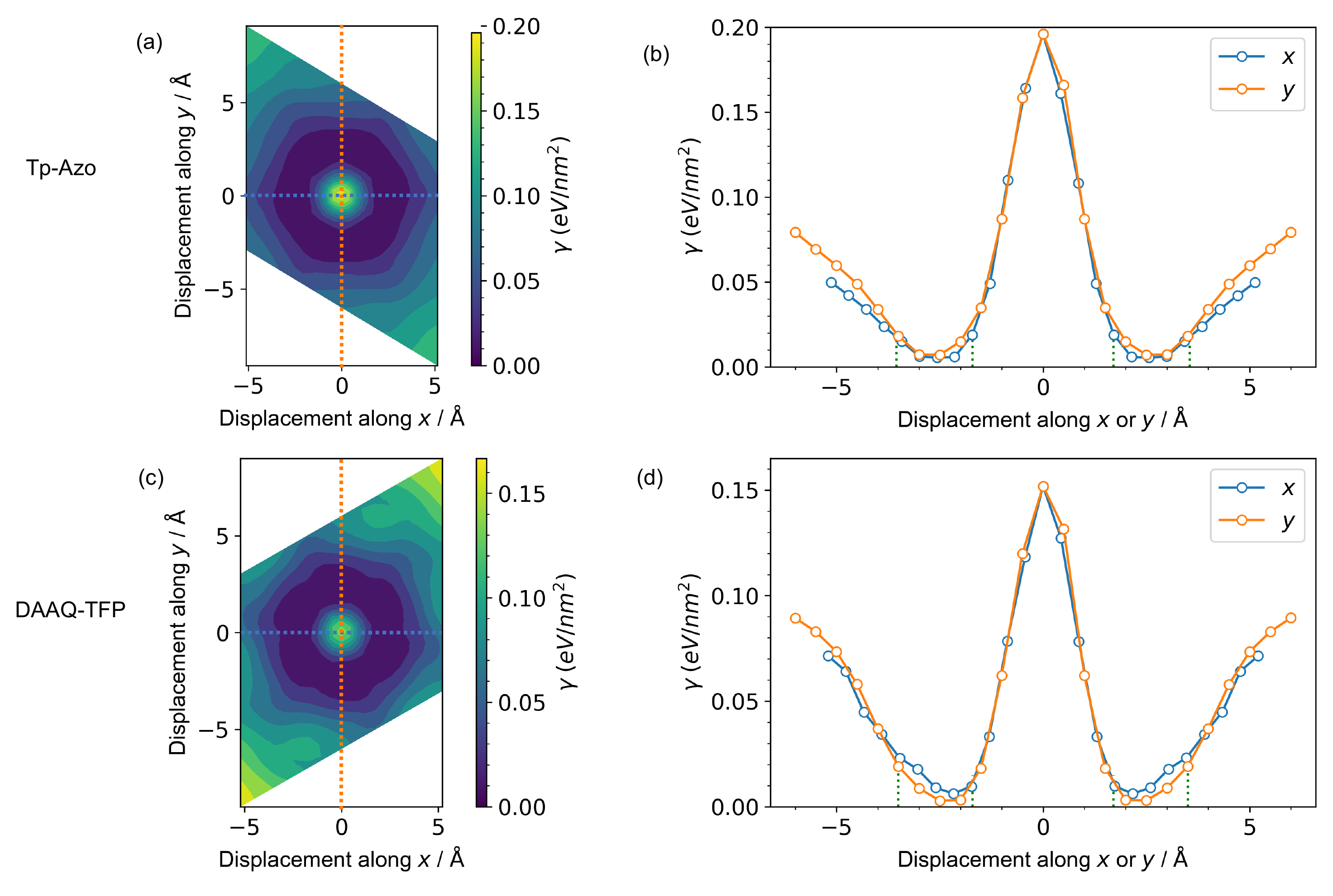}
    \caption{Contour maps of the potential energy surfaces for different displacements along the $a$ and $b$ axes of \textbf{(a)} Tp-Azo and \textbf{(c)} DAAQ-TFP. Blue corresponds to regions of low energy, while yellow regions are high energy. A zero (0 {\AA}, 0 {\AA}) shift at the centre of each plot represents a perfectly eclipsed geometry. \textbf{(b)} and \textbf{(d)} show the 1D cross-sections along the $x$ (blue line) and $y$ (orange line) axes. The blue and orange dashed lines in \textbf{(a)} and \textbf{(c)} correspond to those in \textbf{(b)} and \textbf{(d)}, respectively. The green vertical dotted lines in \textbf{(b)} and \textbf{(d)} are the inner and the outer sides of the low energy rings of \textbf{(a)} and \textbf{(c)}, respectively.}
    \label{fig:3}
\end{figure*}

\begin{figure*}[t!]
    \centering
    \includegraphics[width=0.8\textwidth]{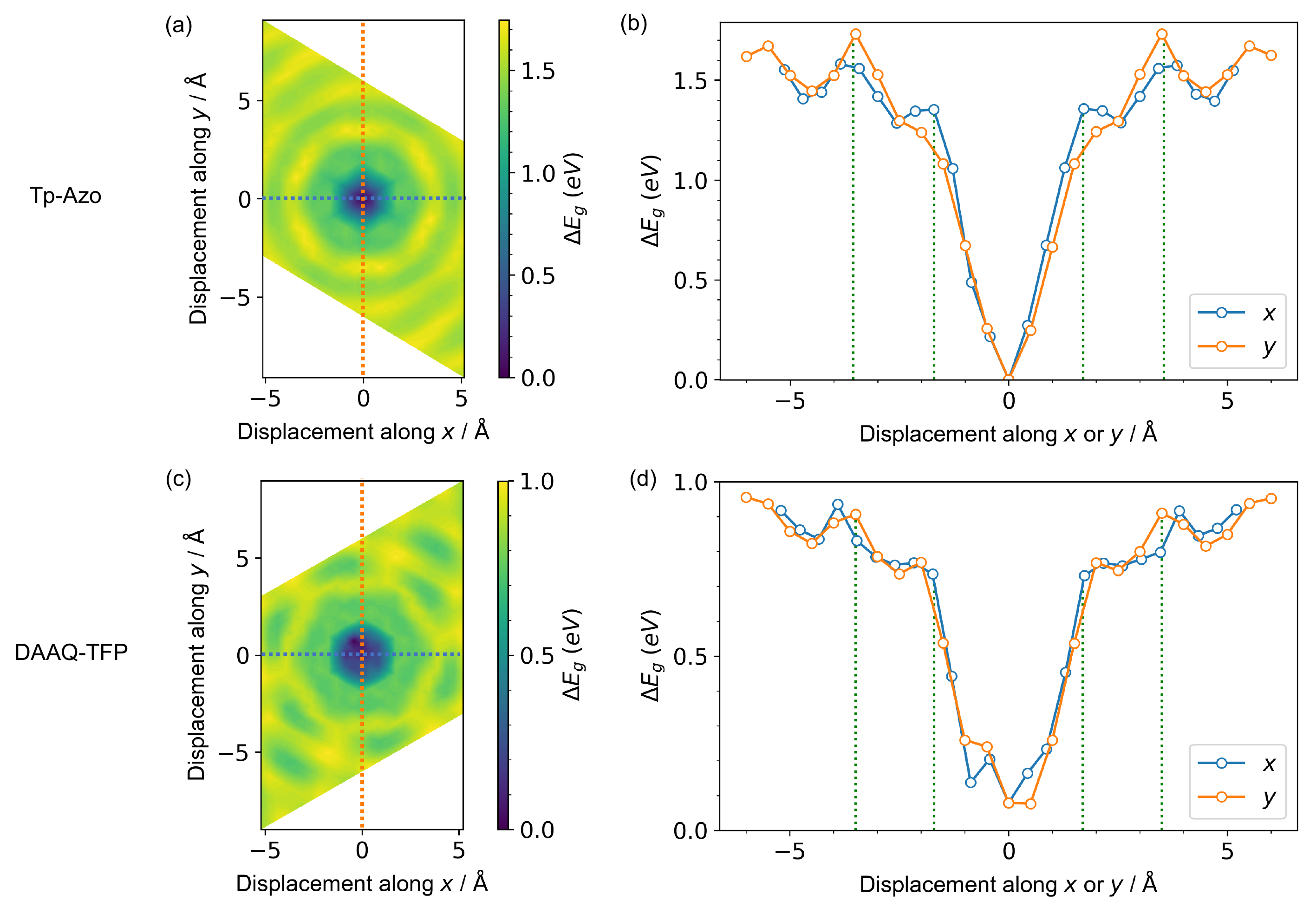}
    \caption{Contour maps of the electronic band gaps for displaced structures of \textbf{(a)} Tp-Azo and \textbf{(c)} DAAQ-TFP, normalized to the minimum band gap and calculated using hybrid DFT (HSE06). Blue / dark green corresponds to small band gaps and yellow represents large band gaps. A zero (0 {\AA}, 0 {\AA}) shift at the centre of each plot represents a perfectly eclipsed geometry. \textbf{(b)} and \textbf{(d)} plot the band gap variation upon displacement along the $x$ or $y$ axes, corresponding to the dashed lines in \textbf{(a)} and \textbf{(c)}, respectively. The green vertical dotted lines in \textbf{(b)} and \textbf{(d)} are the inner and the outer sides of the low energy rings in Fig.~\ref{fig:3}\textbf{(a)} and \textbf{(c)}, respectively.}
    \label{fig:4}
\end{figure*}

\subsection{Binding between layers}
Due to their non-covalent interlayer interactions, the structural properties of COFs can be modified through exfoliation or tuning of interlayer distances.\cite{xu2015stable,wang2017exfoliation} 
Single- or few-layer COFs are an emerging class of functional materials.\cite{li2020partitioning, li2020partitioning,deblase2015rapid} 
For example, nanosheets of DAAQ-TFP show promise in battery cathodes due to shorter ion/electron migration pathways and  higher ionic/electronic diffusion rates.\cite{wang2017exfoliation} 
Hence, knowledge of the binding energy between layers in COFs is important for tuning their performance in device applications. 

The binding between layers of Tp-Azo and DAAQ-TFP was calculated from the total energy difference between the relaxed monolayer and the bulk COFs. Due to the requirement of periodic boundary conditions, the layer distance was increased to 30 \r{A} to ensure negligible chemical interactions between repeating layers (Supplementary Fig.~S2).\cite{bjorkman2012van} The exfoliated COF layers were fully relaxed with this fixed interlayer distance. After relaxation, the undulating monolayer structure became planar again, indicating this distortion to be a result of interlayer interactions (Supplementary Fig.~S2). 
The binding energy, $\gamma$, to form the monolayer can be calculated per unit area according to:
\begin{equation}\label{eq1}
    \gamma = (E_\mathrm{monolayer}-N E_\mathrm{bulk}) / A,
\end{equation}
where $E_\mathrm{monolayer}$ is the total energy of the COF monolayer, $N$ is the total number of atoms in the surface of the monolayer, and  $E_\mathrm{bulk}$ is the bulk energy per atom.\cite{bjorkman2012van, han2019surface} $A$ is the area of the bottom or the top surface of the monolayer. As there is only a single layer per unit cell, only a single layer surface area is needed in Equation \ref{eq1}. The binding energy between layers of Tp-Azo and DAAQ-TFP COFs are 2.5 meV/\r{A}$^2$ and 3.1 meV/\r{A}$^2$, respectively. Compared with the binding energies of other 2D layered materials such as graphite (13 meV/\r{A}$^2$) and \ce{MoS2} (20 meV/\r{A}$^2$)\cite{bjorkman2012van}, Tp-Azo and DAAQ-TFP can be classified as $'$easily exfoliable$'$ 2D materials (specifically, their binding energies are smaller than 30 meV/\r{A}$^2$).\cite{han2019surface, mounet2018two} The high porosity in the COF structure greatly contributes to this low interlayer binding energy, with $\gamma$ increasing to 8.5 meV/\r{A}$^2$ and 9.2 meV/\r{A}$^2$, respectively, when the pores are omitted from the framework surface area $A$ in Equation \ref{eq1}. 

\subsection{Potential energy surface for layer displacements}
A series of displaced structures were generated with the layers offset to varying amounts along the $a$ and $b$ axes (Fig.~\ref{fig:2}). 
For each displaced structure, the internal geometry was relaxed and the energy minimum was set to 0 (Fig.~\ref{fig:3}).
The PES exhibits a characteristic hexagonal shape for both COFs, resembling a ``sombrero'' potential.\cite{koo2012classification, meier2020manifestation}
A similar scan of rigid layers without geometry relaxation is shown in Supplementary Fig. S3; a steeper and more fragmented PES is produced.
The interlayer $\pi$-$\pi$ interactions give rise to a stable hexagonal ring  (dark blue in Fig.~\ref{fig:3}\textbf{(a)} and \textbf{(c)}) where the relative layer displacements maximise the attractive electrostatic interactions.

Eclipsed stacking of layers is significantly less energetically favourable than the displaced arrangements and represents a local maximum on the PES. 
The center of the PES, corresponding to no displacement, is 0.20 $\textrm{eV}/\textrm{nm}^2$ higher than the minimum energy for Tp-Azo, and 0.17 $\textrm{eV}/\textrm{nm}^2$ for DAAQ-TFP. 
This is shown most clearly from the 1D cross-sections in Fig.~\ref{fig:3}\textbf{(b)} and \textbf{(d)}.
The width of the low energy wells is approximately 1.8 \r{A} along both the $x$ and $y$ axes. 

The suggested behaviour is distinct from typical stacking faults associated with discrete local minimum configurations, e.g.~mixtures of hexagonal (AB) and cubic (ABC) packing in close-packed crystals. 
Here, a continuous range of configurations are accessible. 
Random sampling of the low energy ring would produce an average structure that appears as eclipsed to macroscopic measurements,\cite{spitler20112d} yet in reality comprises locally offset COF layers.
Moreover, the soft ``sombrero" PES suggest a high sensitivity of the actual COF structures to the synthesis and processing conditions.   

\subsection{Electronic band gap opening}
Next, we consider the impact of these displacive instabilities on the underlying electronic structure of the COFs. 
Remarkably, the band gap variation follows the inverse of the PES.
The smallest band gap is exhibited by the eclipsed structure with no displacements along the $a$ and $b$ axes (at the center of the heatmaps). 
The HSE06 calculated band gap is 0.28 eV for eclipsed Tp-Azo and 1.29 eV for eclipsed DAAQ-TFP. 
A band gap opening of 1.37 eV (to 1.65 eV) and 0.75 eV (to 2.04 eV) is found for displaced Tp-Azo and DAAQ-TFP, respectively. 
A similar behaviour has previously been observed in COF-5.\cite{kuc2020proximity}
We note that monolayers of Tp-Azo and DAAQ-TFP exhibit even larger band gaps of 2.06 eV and 2.36 eV as a result of quantum confinement in the 2D sheets (Supplementary Fig.~S4).

These band gaps suggest that Tp-Azo and DAAQ-TFP COFs are semiconducting materials. 
Fig.~\ref{fig:4}\textbf{(b)} and \textbf{(d)} show that the band gaps change sharply upon small displacement between layers within a deep well of 2 \r{A} width. 
However, when the displacement is more than 2 \r{A} but less than 6 \r{A}, the band gap  oscillates between 1.25 eV and 1.72 eV for Tp-Azo, and 0.71 eV and 0.97 eV for DAAQ-TFP. 
Our analysis suggests a relatively small variation within the low energy ring that should be populated at room-temperature in thermal equilibrium. 
\hl{The magnitude of the band gap plays an important role in battery applications. It is connected to the open-circuit voltage set by the electrochemical potentials of the anode and cathode, and also connects to charge transport (electrodes must conduct ions and electrons) as well as the stability windows.\cite{goodenough2010challenges,cherkashinin2019performance} } 

\begin{figure*}[t!]
    \centering
    \includegraphics[width=\textwidth]{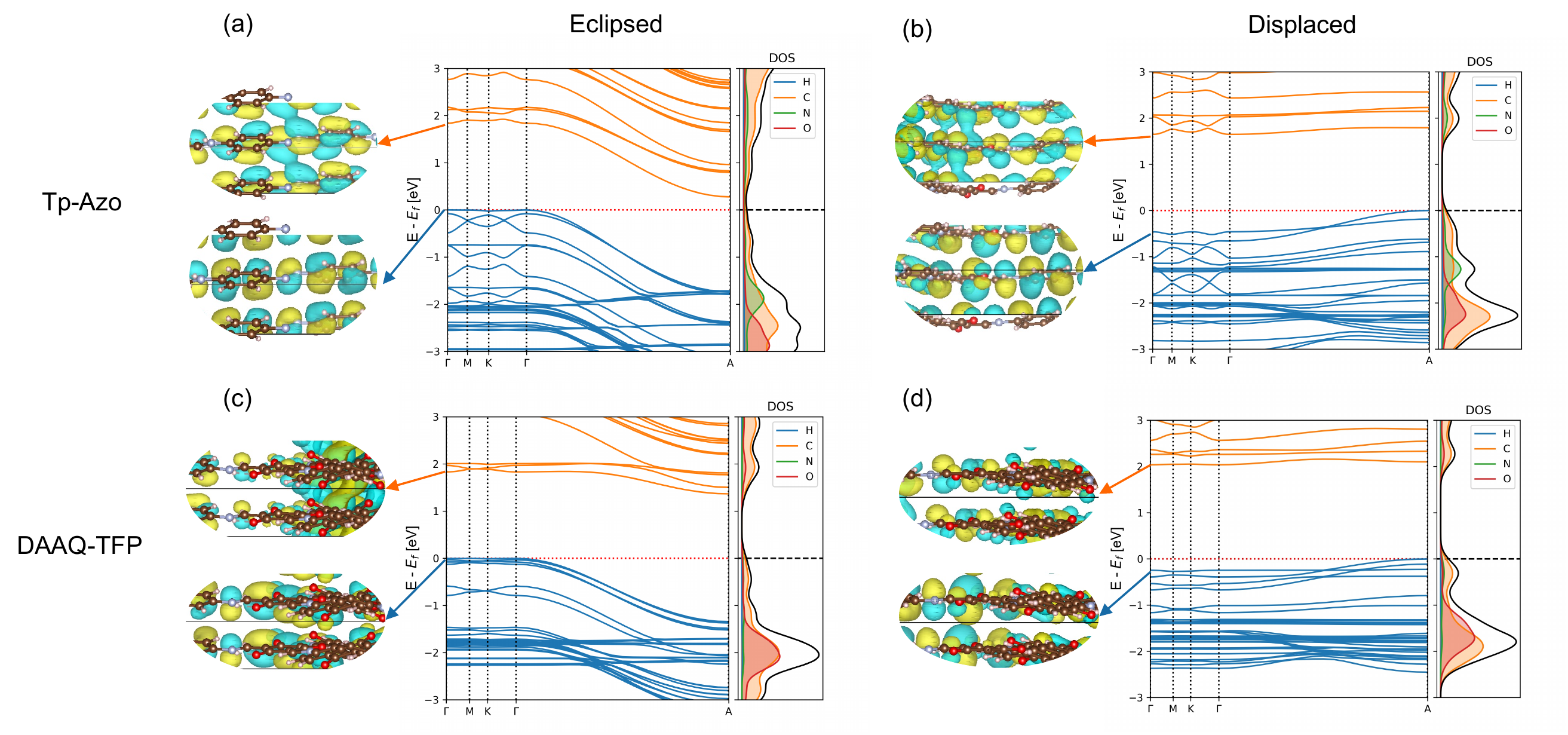}
    \caption{Electronic band structures of eclipsed and displaced Tp-Azo \textbf{(a-b)} and DAAQ-TFP \textbf{(c-d)}, alongside the electronic wavefunctions (isosurface = 0.004 eV/\r{A}$^3$) of the valence band maximum and conduction band minimum at the $\Gamma$ point. The highest occupied band is indicated by the dashed horizontal line.}
    \label{fig:5}
\end{figure*}

\subsection{Origins of strong electronic coupling to layer displacements}
The electronic band structures of the eclipsed and slipped Tp-Azo and DAAQ-TFP COFs are compared in Fig.~\ref{fig:5}. 
Both COFs exhibit low band dispersion along the $\Gamma$-M-K-$\Gamma$ path in reciprocal space, which corresponds to in-plane directions. 
The layer stacking direction, which is the shortest axis in real space, corresponds to the longer $\Gamma$-A line in the band structure. 

For eclipsed stacking, the interlayer interactions produce dispersive bands with a band width 1.72 eV in the upper valence band (VB) and 1.56 eV in lower conduction band (CB) along the $\Gamma$-A path of Tp-Azo. The dispersion of DAAQ-TFP is slightly reduced, giving rise to a band width of to 1.35 eV in the VB and 0.47 eV in the CB in the interlayer direction. 
Eclipsed Tp-Azo and DAAQ-TFP both have strongly indirect band gaps arising from the interlayer interactions between between $\Gamma$ and A. 
 
Layer slippage results in a pronounced change in the band dispersion.
The band structures remain weakly indirect between $\Gamma$ and A, but the dispersion itself is inverted with the VB maximum changing location. 
Both the conduction and valence bands become much flatter upon layer displacement, particularly along the $\Gamma$-A path. In going from the eclipsed to displaced stacking mode, the widths of the topmost valence bands reduce from 1.72 to 0.53 eV (Tp-Azo) and from 1.35 to 0.28 eV (DAAQ-TFP), and from 1.65 to 0.15 eV (Tp-Azo) and 0.53 to 0.13 eV (DAAQ-TFP) for the bottom-most conduction bands.

The corresponding $\Gamma$ point wavefunctions are shown in the insets of Fig.~\ref{fig:5}.
They confirm that the band edges are formed from the C 2p$_z$ $\pi$ subsystem.
For the eclipsed structure, the interlayer interactions are strongly anti-bonding at the $\Gamma$ point.
This explains the strong downward dispersion towards A, where the phase of successive layers is reversed.  
In the displaced structure, stronger interlayer $\pi$ bonding interactions are allowed at the $\Gamma$ point and the band dispersion is suppressed along the $\Gamma$-A line.
These changes result in a band gap that is weakly indirect and much larger in magnitude compared to the eclipsed structure.  

\section{Conclusions}
It is convenient to represent and model covalent organic frameworks as an ordered sequence of eclipsed planar layers.
However, by taking the examples of Tp-Azo and DAAQ-TFP, we have shown that the results can be misleading in line with recent observations for other layered COFs.
A displaced stacking sequence of undulating layers both lowers the total energy of the frameworks and results in a large change in the electronic structure driven by interlayer $\pi$ orbital overlap.
Layer displacements produce a pronounced band gap opening in these frameworks.

The unusual ``sombrero'' potential energy surface for layer displacements, which mirrors the variation in band gap, has important implications.
Although macroscopically a given COF may appear to have an eclipsed structure, for example on the basis of diffraction measurements, locally a continuous range of stacking sequences are accessible. 
The strong coupling between layer orientation and electronic structure highlights the potential for COF twistronics where longer range modulations in the crystal potential are harnessed. 
\hl{Layer offsets may be controlled by various experimental approaches, such as chemical intercalation, synthetic modification of composition including aromatic ring size, and processing temperature.\cite{emmerling2021interlayer,putz2020total}}

These findings will be of particular importance when screening COFs for applications in energy storage and conversion where electrochemical and photochemical descriptors are significantly altered including accessible voltage ranges for batteries, stability windows for electrocatalysis, and visible light absorption for photoelectrochemical systems. 

\section*{Author Contributions}
Ju Huang: Conceptualization, Data curation, Formal analysis, Investigation, Methodology, Visualization, Writing – original draft, Writing – review \& editing.
Matthias J. Golomb: Methodology, Formal Analysis, Software, Writing – review \& editing.
Seán R. Kavanagh: Software, Writing – review \& editing.
Kasper Tolborg: Formal Analysis, Visualization, Software, Writing – review \& editing.
Alex M. Ganose: Visualization, Software, Writing – review \& editing.
Aron Walsh: Conceptualization, Methodology, Resources, Supervision, Formal Analysis, Writing – review \& editing.

\section*{Conflicts of interest}
The authors have no conflicts of interest to declare. 

\section*{Acknowledgements}
J.H.~thanks Chengcheng Xiao for suggestions relating to the computational workflow.
J.H.~acknowledges Imperial College London and the Chinese Scholarship Council (CSC) for providing a PhD scholarship. 
S.R.K.~acknowledges the EPSRC Centre for Doctoral Training in the Advanced Characterisation of Materials (CDT-ACM)(EP/S023259/1) for funding a PhD studentship. K.T.~acknowledges the Independent Research Fund Denmark for funding through the International Postdoctoral grant (0164-00015B).
A.M.G~was supported by EPSRC Fellowship EP/T033231/1.
We are also grateful to the UK Materials and Molecular Modelling Hub for computational resources, which is partially funded by EPSRC (EP/P020194/1 and EP/T022213/1). Via our membership of the UK's HEC Materials Chemistry Consortium, which is funded by EPSRC (EP/L000202), this work used the ARCHER2 UK National Supercomputing Service (https://www.archer2.ac.uk). 



\balance


\bibliography{ref} 

\bibliographystyle{rsc} 

\end{document}


\beginsupplement

\captionsetup[figure]{labelfont={bf},labelformat={default},labelsep=period,name={Fig.}}
\captionsetup[table]{labelfont={bf},labelformat={default},labelsep=period,name={Tab.}}
%
%

\begin{table}[hbt!]
\caption{Lattice parameters and space groups of \textbf{(a)} planar experimentally-reported crystal structure and \textbf{(b)} wavy relaxed structure of Tp-Azo. The lattice parameters and space groups of \textbf{(c)} planar experimentally-reported crystal structure and \textbf{(d)} wavy relaxed structure of DAAQ-TFP.}
\begin{center}
\begin{tabular}{lccccccc}
\hline
\multicolumn{8}{|c|}{Lattice parameters} \\
\hline
-- & $a$ (\r{A}) & $b$ (\r{A}) & $c$ (\r{A}) & $\alpha$ ($^{\circ}$) & $\beta$ ($^{\circ}$) & $\gamma$ ($^{\circ}$) & Space group\\
\hline
\textbf{(a)} Tp-Azo\cite{chandra2014phosphoric} &31.50 & 31.50 & 3.30  & 90 & 90  & 120 & $P6/m$\\

\textbf{(b)} Tp-Azo$^*$ &33.28 & 33.36 & 4.23  & 61.71 & 128.44  & 121.17 & P$\bar{1}$\\

\textbf{(c)} DAAQ-TFP\cite{deblase2013beta} & 29.83 & 29.83 & 3.60  & 90 & 90  & 60 & $P6/m$\\

\textbf{(d)} DAAQ-TFP $^*$ &30.28 & 30.38 & 4.04  & 68.68 & 56.15  & 60.19 & P$\bar{1}$\\

\hline  
\end{tabular}
\end{center}
\label{tab1:lattice_parameter}
\end{table}

\begin{figure*}[t!]
    \centering
    \includegraphics[width=0.8\textwidth]{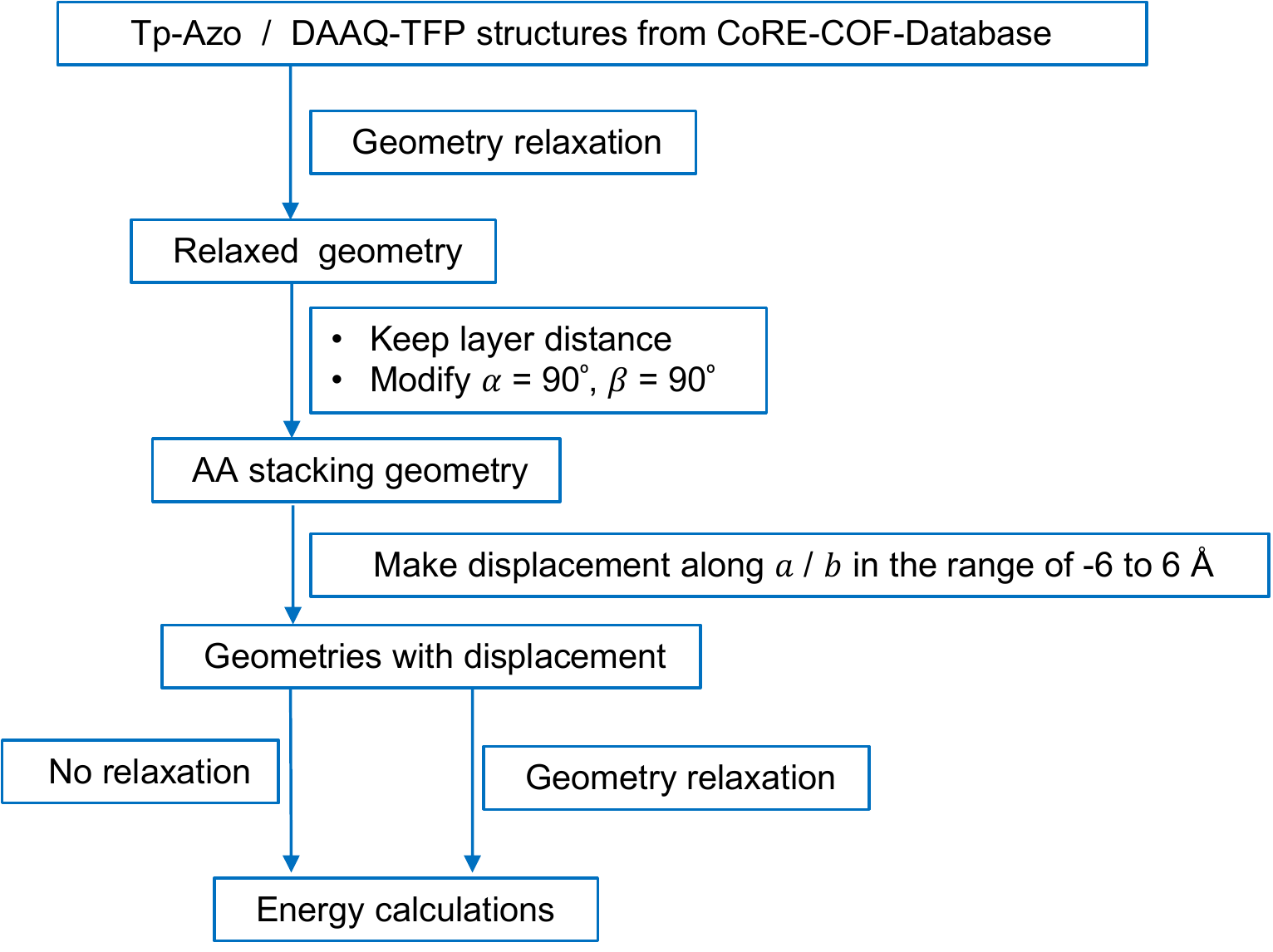}
    \caption{Workflow of displacement process.}
    \label{fig:1}
\end{figure*} 

\begin{figure*}[t!]
    \centering
    \includegraphics[width=\textwidth]{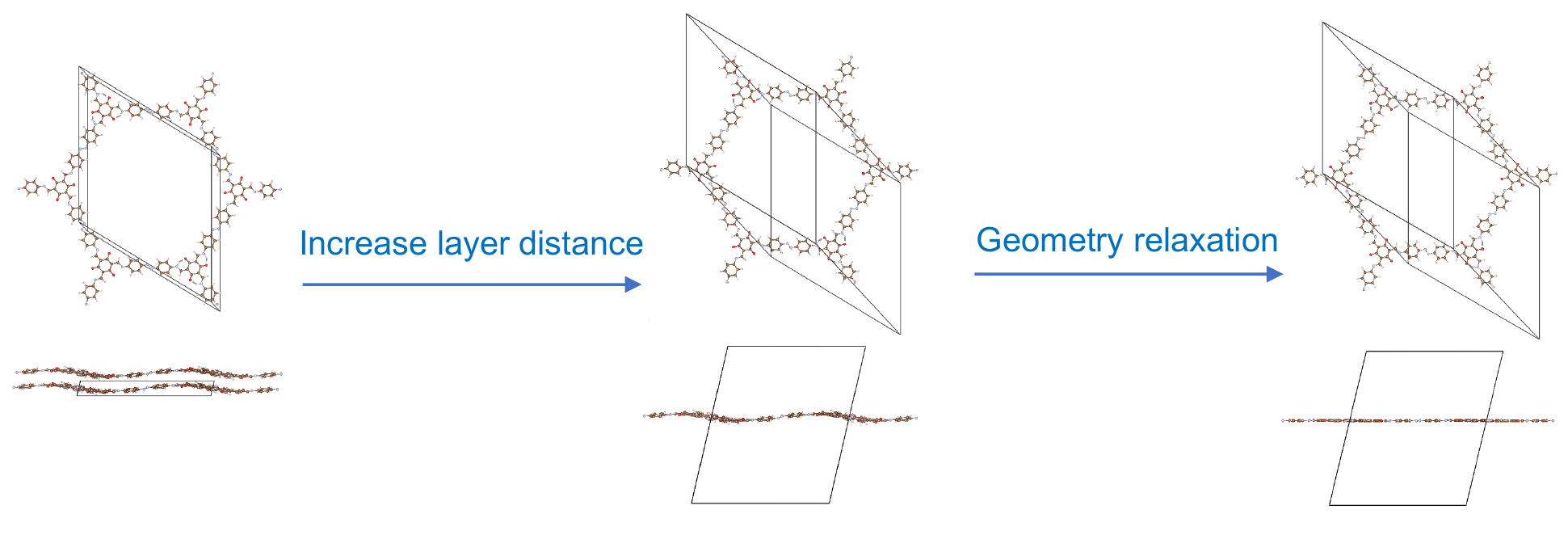}
    \caption{Process of forming monolayer and the binding energy calculation of Tp-Azo COF. The upper figures are viewed looking down the $c$ axis, and the lower figures are viewed along the $ab$ plane. The layer distance in geometry relaxation keeps 30 \r{A}.}
    \label{fig:2}
\end{figure*} 



\begin{figure*}[t!]
    \centering
    \includegraphics[width=\textwidth]{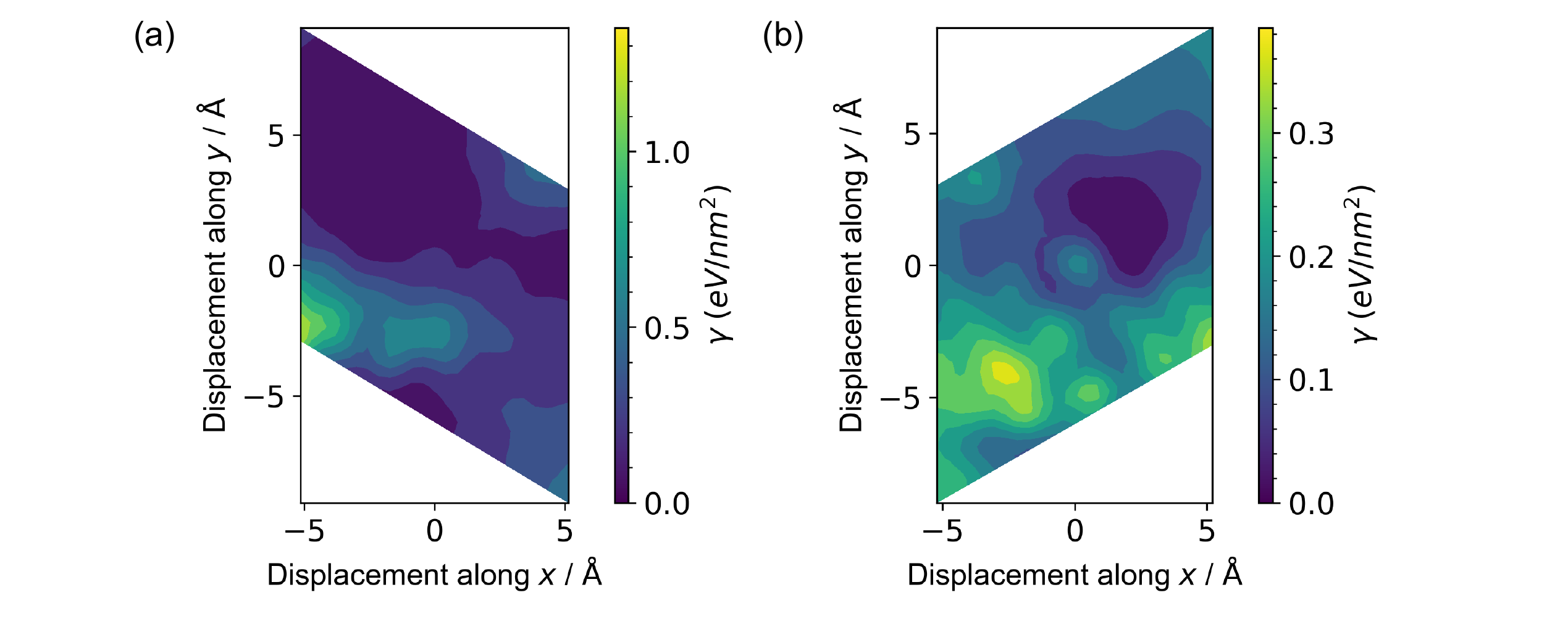}
    \caption{Potential energy surfaces (PES) for rigid layer displacements along $a$ and $b$ sides of (a) Tp-Azo and (b) DAAQ-TFP. Note, the PES reported in the main text includes geometry relaxation.}
    \label{fig:3}
\end{figure*} 

\begin{figure*}[t!]
    \centering
    \includegraphics[width=\textwidth]{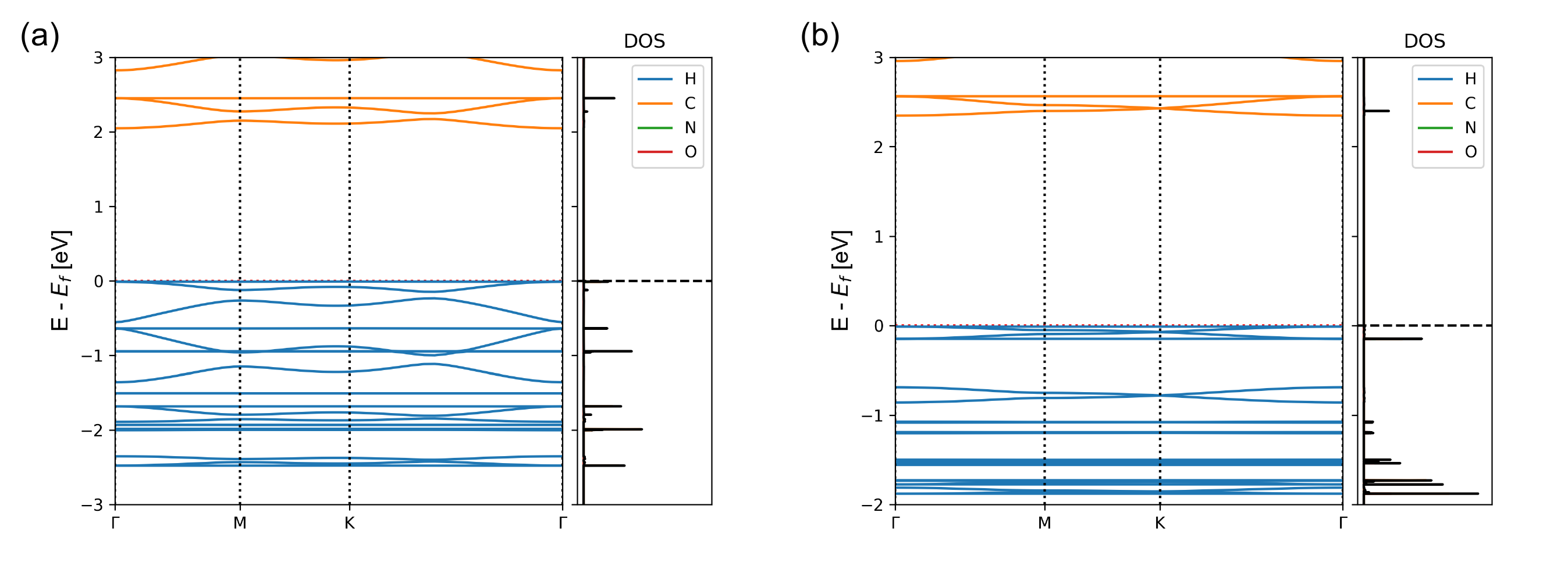}
    \caption{Electronic band structures of monolayer Tp-Azo (a) and monolayer DAAQ-TFP (b).}
    \label{fig:4}
\end{figure*}

\clearpage
\bibliography{ref.bib}
\endsupplement